# Nonexistence of Chaotic Solutions of Nonlinear Differential Equations

Lun-Shin Yao
School of Mechanical, Aerospace, Chemical and Materials Engineering
Arizona State University
Tempe, AZ 85287
USA

## *Abstract*

We discuss some important issues arising from computational efforts in dynamical systems and fluid dynamics. Various individuals have misunderstood these issues since the onset of these problem areas; indeed, they have been routinely misinterpreted, and even viewed as "laws" by some. This paper hopes to stimulate appropriate corrections and to realign thinking, with the overall goal being sound future progress in dynamical systems and fluid dynamics.

## *I. Introduction*

Both existence and uniqueness require consideration when studying a system of differential equations. The existence of chaotic solutions for differential equations (not discrete algebraic maps) has never been proven; instead, existence is routinely taken for granted. However, differential equations presumed to be chaotic are *unstable*; therefore, no *convergent* numerical solutions for them are possible, as clearly pointed out by Von Neumann. It is likely that all published numerical solutions are simply numerical errors [Yao, 2007; Yao, 2010; Yao & Hughes, 2008a, 2008b]. The upshot of this conclusion is that there exists the frightening possibility that no solutions of chaotic differential equations are obtainable through any existing discretized numerical method, implying that solutions *do not exist*. Yao [2010] discussed this matter in detail for the Lorenz system. Without a resolution of the existence question, it seems frivolous to discuss properties of possibly non-existing solutions of differential equations. In fact, the relevance of Smale's horseshoe to differential equations is questionable. We discuss these matters in the first two open problems defined below.

The relevance of chaos to turbulence is an unsettled question. Chaos is random in time; turbulence is random in time and space. We believe they share some fundamental properties. On the other hand, multiple solutions of the Navier-Stokes equations exist when the Reynolds number is larger than its critical value; they reflect sensitivity *to initial conditions* [Yao, 2009]. The selection principles for an appropriate solution and the resulting nonlinear wave interactions for different initial conditions are explained by the resonant wave theory developed by Yao [1999]. However, multiple solutions of the Navier-Stokes equations evolve in an entirely different manner than the description of Smale's horseshoe; this leads to the third open problem discussed below. It also provides evidence that sensitivity to initial conditions cannot be a sufficient condition for chaos, as many practitioners of dynamical systems believe.

## *2. Open Problems.*

**Open Problem 1. No chaotic solution of a system of nonlinear differential equations exists.**

The Poincare-Bendixson Theorem clearly implies that a two-dimensional continuous dynamical system cannot give rise to a strange attractor. Some published chaotic solutions for a single nonlinear or linear ordinary differential equation violate this theorem. They are unstable *spurious* numerical solutions, but are frequently misinterpreted as chaos [Griffiths, Sweby & Yee, 1992]. Since no analytical chaotic solution is available for a system of more than two nonlinear differential equations, a numerical solution becomes the only alternative. Systems of ordinary differential equations that exhibit chaotic responses have yet to be correctly integrated. So far no *convergent* computational results have been confirmed for chaotic differential equations. Various computed numbers are not solutions of the continuous differential equations; all proposed chaotic responses are, more likely than not, simply numerical noise and have nothing to do with the solutions of differential equations. Two well-known Lorenz models [Lorenz, 1963, 1990] are typical examples showing that computed non-periodic solutions are simply due to the unstable amplification of truncation errors and the distribution of singularities [Teixeira, Reynolds & Judd, 2007; Yao & Hughes, 2008b; Yao, 2010].

Numerical methods convert continuous differential equations to a set of algebraic equations to be solved by computers. Derivatives in the continuous equations are replaced by corresponding discretized forms. For example, $\frac{dx}{dt} \approx \frac{\Delta x}{\Delta t}$, where $\frac{\Delta x}{\Delta t}$ equals $\frac{dx}{dt}$ in the limit as $\Delta t$ approaches zero. Instead of taking the limit, $\Delta t$ has a finite "small" value in the discretizing procedure such that $\frac{\Delta x}{\Delta t}$ sufficiently approximates $\frac{dx}{dt}$ when $\Delta t$ is "small." Von Neumann established that discretized algebraic equations must be *consistent* with the differential equations, and must be *stable* in order to obtain *convergent* numerical solutions for the given differential equations. This can be easily checked by ensuring that the difference between computed results for successively reduced time-step size is acceptably small. A successful check can be taken as an indication that $\frac{\Delta x}{\Delta t}$ is indeed a good approximation to $\frac{dx}{dt}$ and that the solutions of the discrete approximations have *converged* to the solution of the continuous equations. *Any computed results that fail this check are simply numerical errors that have no mathematical meaning.* This is the most important fundamental rule in solving differential equations numerically by computers. Unfortunately, this rule is consistently ignored in computational chaos and turbulence, and researchers in numerical chaos have often argued that it is irrelevant. One wonders how a set of discretized algebraic equations can be related to differential equations if convergence cannot be assured.

Yao and Hughes [2008b] repeated the Lorenz computation [2006a] and found that his numerical solutions do not converge. The positive Liapinov exponents found in his computation are simply due to unstable computations, which violate Von Neumann's convergence criteria. It is worthwhile to note that Lorenz's second model [Lorenz, 1990] is not uniformly dissipative and does not have an attractor, but its trajectory settles inside a large attracting set. The unstable computation of Lorenz's first model [Lorenz, 1963] has been analyzed and discussed in detail by Yao [2010]. Some important misconceptions will be discussed below in Open Problem 2. Additional examples of unstable computation of nonlinear differential equations have also been discussed and can be found in [Yao, 2010]. Comprehensive and carefully computed examples that show useless computational results, which are integration time-step dependent, can be found in [Teixeira, Reynolds & Judd, 2007; Yao & Hughes, 2008a].

It would be an exciting contribution if a convergent computed chaotic solution for any system of nonlinear differential equations could be obtained. According to our experience, this would require a breakthrough in computational methods. A much larger and faster computer alone will not solve this problem. Beyond that, there is no reason to believe that a chaotic or turbulent solution exists for differential equations, including the Navier-Stokes equations in fluid mechanics.

**Open Problem 2. Is Smale's horseshoe relevant to chaotic solution of differential equations, including direct simulation of turbulence.**

Smale's horseshoe lays the foundation of the structural stability of dynamic systems, chaos. The studies in the sixties concluded that a hyperbolic dynamic system is structurally stable. In a hyperbolic system, the tangent space at every point can be decomposed into three complementary directions, which are stable, unstable and neutral. However, no system of differential equations is hyperbolic [Viana, 2000].

In the 14th of 18 challenging mathematical problems [Smale, 1998], Smale asked: Is the dynamics of the ordinary differential equations of Lorenz [1963] that of the geometric Lorenz attractor of Williams, Guckenheimer and Yorke? One can view this as a general question about the equivalence of discrete algebraic maps and differential equations. In spite of numerous attempts, a convincing proof of the existence of the geometric Lorenz attractor for the Lorenz's differential equations was not achieved until Tucker [2002] provided a solution to this problem with the aid of his computer. He showed that:

1. The Lorenz equations are dissipative; consequently, they contain a forward invariant region. This shows it is a *closed invariant* set.

2. The result is sensitive to initial conditions so it is a strange attractor.

3. The trajectory is indecomposable and *looks* topologically transitive.

On the other hand, Yao [2010] reported that computed chaotic solutions of the Lorenz equations are consistently contaminated by the amplification of truncation errors, and are sensitive to artificial integration time-steps. He carried out a thorough analysis of the first Lorenz model and showed that two mechanisms can amplify truncation errors. One error is well known; it is amplified along the unstable manifold and causes an *exponential* amplification of truncation errors. This violates the stability requirement of Von Neumann's convergence so the computational results have no value. However, if one follows Tucker's method, it can be shown that the numerical results, which are dependent on the integration time-steps, satisfy all three properties of Lorenz algebraic maps!

The second mechanism is the explosion amplification of truncation errors. This is because the trajectory penetrates the virtual separatrix, and violates the differential equations. The existence of a virtual separatrix is a consequence of singular points of a non-hyperbolic system of differential equations, which is not *shadowable*.

Sensitivity to initial conditions has been studied for hyperbolic systems [Bowen, 1975], and is an active topic for nearly hyperbolic systems [Dawson, Grebogi, Sauer, & Yorke, 1994; Viana, 2000; Palis, 2000; Morales, Pacifico & Pujals, 2002]. In particular, the methods of shadowing for hyperbolic systems have shown that trajectories may be *locally* sensitive to initial conditions while being *globally* insensitive since true trajectories with adjusted initial conditions exist. Such trajectories are called shadowing trajectories, and lie very close to the long-time computed trajectories. However, systems of differential equations arising from physical applications are not hyperbolic systems. If the attractor is *transitive* (ergodic), all trajectories are inside the attractor so they are generally *believed* to be solutions of the underlying differential equations, whatever the initial conditions. Little is known about systems, which are not transitive (non-ergodic). Do & Lai [2004] have provided a comprehensive review of previous work, and discussed the fundamental dynamical process indicating that a long-time shadowing of non-hyperbolic systems is not possible.

Another commonly cited computational example in chaos involves two solutions of slightly different initial conditions that remain "close" for some time interval and then diverge abruptly. In fact, this behavior is often believed to be a typical characteristic of chaos. More properly, this phenomenon is actually due to the explosive amplification of numerical errors, and violation of the differential equations as described above. Similar computational results can occur for two different integration time steps; many would consider such results as acceptable since it is a "twin brother" of sensitivity to initial conditions and a typical characteristic of chaos. This mistake deserves clarification.

In fact, two mechanisms of error amplification can alter the topological properties of attractors. This shows the common expectation that the presence of an attractor guarantees the correctness of computed results is completely wrong. *It is easy to conclude that Tucker did not prove that the Lorenz system has the property of a Lorenz map since a solution of the Lorenz system does not exist.*

Before we move to the next open problem, we would like to point out that the inviscid-flow equations do not have non-periodic turbulent solutions. This is because these equations are not dissipative. *Those studying chaos in conservative systems, which cannot have attractors, should check their computations to ensure that their results are not dependent on the integration time-step.*

**Open Problem 3. What is the relation between Smale's horseshoe and the physics of the Navier-Stokes equations?**

That the structure of *resonances* determines the long-time solutions of *nonlinear* differential equations was originally advocated by Stokes [1847] and Poincare [1892]; the later developed the well-known Poincare-Lighthill-Kuo (PLK) Method. Yao [1999] proposed an eigenfunction-spectral method, a fully nonlinear approach and a generalization of the PLK method, for a direct numerical solution of the Navier-Stokes equations. In the following discussion, the generalized eigenfunctions are referred to as "waves." The linear terms of the Navier-Stokes equations involve the local acceleration terms, the viscous dissipative terms, and the pressure-gradient terms. The nonlinear terms represent the energy transfer among all waves, traditionally referred to as inertial effects. There are two kinds of transfer: forced transfer and resonance transfer. The forced transfer can create a new wave, but with limited energy transfer. Substantial energy transfer can only occur when the resonance conditions are satisfied [Phillips, 1960].

The basic idea of the eigenfunction-spectral method is that a solution of nonlinear differential equations can be expressed in terms of a complete set of waves. The amplitude of every wave is not necessarily nonzero. The base flow is a special wave, which has zero frequency and infinite wavelength. The pressure gradient drives the base flow, and is a source of energy. For the Reynolds number, Re, less than its critical value, all disturbance waves are stable and highly dissipative; thus, only the base flow exists. As Re reaches its critical value, the energy received by the critical wave from the base flow balances with dissipation so its amplitude does not change with respect to time. This is the condition of neutral stability, or the onset of instability. Any further increase of Re, the energy from the base flow becomes larger than that can be dissipated by waves; the amplitude of the waves will grow. As Re increases further, more waves become unstable and the nonlinear wave interaction becomes complicated. This is usually called finite-amplitude instability. We found that convergent computation becomes more difficult, and computations eventually diverge when Re increases further, but is still far below its transitional value for the flow to become turbulent. This fact is not well known and respected; many researchers routinely compute finite-amplitude instabilities and turbulence, generating numerical errors, without bothering to check their convergence.

The nonlinear interactions among Taylor-Couette vortices (driven by boundary motion) with different wave numbers were studied by using a weakly nonlinear theory [Yao & Ghosh Moulic, 1995a] and nonlinear theory [Yao & Ghosh Moulic, 1995b]. Both theories can be derived from the eigenfunction-spectral method as truncated solutions. They represent the disturbance by a Fourier integral and derive an integro-differential equation for the evolution of the amplitude density function of a continuous spectrum. Their formulation allows nonlinear energy transfer among *all* participating waves and remedies the shortcomings of equations of the Landau-Stuart type. Numerical integrations of this integro-differential equation indicate that the equilibrium state of the flow depends on the wave number and amplitude of the initial disturbance as observed experimentally and cannot be determined uniquely on each stable bifurcation branch without knowing its history. The accessible wave numbers of the nonlinear stable range lie within the linear unstable range minus a small band, which was traditionally considered as the range of side-band instability. The numerical results simply indicate that waves in this small band lose more energy than they gain; therefore, waves within this small band decay and excite their harmonics. This agrees with Snyder's observation [Snyder, 1969] that *the trend of energy transfer is determined by the stability characteristics of the base flow.* It is a consequence of fully nonlinear wave interactions, and is not due to *sideband instability* as speculated by weakly nonlinear theories; this misconception has prevailed for a long time, and is still pursued by some classical fluid dynamists.

The analysis of mixed convection in a vertical annulus using a nonlinear theory with continuous spectrum [Yao & Ghosh Moulic, 1994, 1995b] yields a conclusion identical to that from the study of Taylor-Couette flows. They treated the spatial problem as a temporal problem. This is possible by following a control-mass system. Plots of the evolution of the kinetic-energy spectra for the two problems are identical in shape and differ only in amplitudes and ranges of wave numbers. This indicates that the evolution of energy spectra provides *universal* and *fundamental* information for all flows. The selection of the equilibrium wave number is a result of nonlinear wave interactions.

The principles of nonlinear interaction of participating waves are universal and fundamental, and are shared by all flows. The mechanism that allows multiple solutions to exist and the required associated conditions have been demonstrated. The *selection* principles, due to its property of *sensitive to initial conditions,* deduced from the numerical study of [Ghosh Moulic & Yao, 1996; Yao, 1999; Yao& Ghosh Moulic, 1995a; 1995b], are listed below:

1) When the initial disturbance consists of a single dominant wave within the nonlinear stable region, the initial wave remains dominant in the final equilibrium state. Consequently, for a slowly starting flow, the critical wave is likely to be dominant.

2) When the initial condition consists of two waves with finite amplitudes in the nonlinear stable region, the final dominant wave is the one with the larger initial amplitude. If the two waves have the same initial finite amplitude, the dominant wave will be the one closer to the critical wave. On the other hand, if the initial amplitudes are very small, the faster growing wave becomes dominant.

3) When the initial disturbance has a uniform broadband spectrum of noise, the final dominant wave is the fastest linearly growing wave, if the initial amplitude is small. On the other hand, if the uniform noise level is not small, the critical wave is the dominant equilibrium wave.

4) Any initial disturbance outside the accessible frequency range will excite its sub-harmonics or super-harmonics, dependent on the existence of broadband noise, whichever are inside the accessible frequency range. The accessible frequency range is the linear stability range minus a small band.

Similar principles of nonlinear energy transfer have also been found in other simple nonlinear partial differential equations [Yao, 2007, 2009]; thus, the discussion outlined above is general for all nonlinear differential equations. *The open question is: Does a one-to-one correspondence exist between resonance and Smale's horseshoe?*

From the above discussion, one can easily conclude that a sensitivity-to-initial condition is not sufficient for the existence of a strange attractor since such conditions are also true for all instabilities when the Re is larger than its critical value. A strange attractor may exist only if the dynamics have a closed invariant set, and it is sensitive to initial conditions. On the other hand, Open Problem 1 demonstrates that the existence of a strange attractor does not guarantee the correctness of the computed solution. The fact is that many numerical generated attractors are simply the consequence of amplification of truncation errors associated with discrete numerical methods.

Another interesting, but not well known example showing the importance of nonlinearity of the Navier-Stokes equations is related to the calculation of the critical Reynolds number. The critical Re of linear-stability theory (the linear-stability limit) is the value of Re above which ever-present *single* infinitesimal critical wave grows. Equivalently, a critical Re can be defined when the initial amplitude of the critical wave does not change with time as described above. The linear-stability theory is a single-wave theory. The physics of flow instability agrees with the nonlinear theory involving many waves, and differs from the linear theory. The critical Re, calculated by the nonlinear theory, could be a function of the initial amplitude of the critical wave, since energy can be transferred to harmonics and sub-harmonics of the critical wave. Without considering nonlinear effects, all disturbances at *the critical Re, calculated by a linear-stability analysis, will decay to zero* [Yao, 2007].

It is worth noting that conventional studies of fluid mechanics often relied on flow visualization and simple analyses, which can only provide limited information about a particular base flow. In particular, the details of the eigenfunctions discussed earlier, which are important to an understanding of the stability characteristics of a base flow, cannot be total revealed by relatively crude visualization methods and simple analyses. Thus, in the study of certain flows, early researchers may have concluded their work without "seeing" all important features of the flow, causing them to overlook the existence of other possible solutions. The principle of multiple solutions is an important aspect of fluid mechanics that has been largely overlooked [Yao, 2009]. This character of multiple solutions is shared by all nonlinear differential equations as a consequence that their *solutions are sensitive to initial conditions*.

## 3. References:

Bowen, R. [1975] "Equilibrium states and the ergodic theory of Anosov diffeomorphism," 470, *Lecture Notes in Mathematics*, Spring-Verlag

Dawson S., Grebogi C., Sauer T. & Yorke J. A. [1994] "Obstructions to shadowing when a Lyapunov exponent fluctuates about zero," *Physical Review Letters*, 73, 1927-1930.

Do Y. & Lai Y. C. [2004] "Statistics of shadowing time in non-hyperbolic chaotic systems with unstable dimension variability," *Physical Review E,* 69, 016213.

Sandipan Ghosh Moulic, and L.S. Yao [1996] "Taylor-Couette instability of traveling waves with a continuous spectrum," *J. Fluid Mech.*, **324**, 181-198.

Griffiths, D. F., Sweby, P. K. & Yee, H. C [1992] **"**On spurious asymptotic numerical solutions of explicit Runge-Kutta methods," IMA Journal of Numerical Analysis, 12(3):319-338.

Lorenz, E. N. [1963] "Deterministic Nonperiodic Flow," *Journal of the Atmospheric Sciences*, **20**, 130-141.

Lorenz, E. N. [1990] "Can Chaos and Intransitivity lead to Interannual Variability?" *Tellus* **42A**, 378-389.

Lorenz, E. N.[2006a] "Computational periodicity as observed in a simple system," *Tellus*, **58A**, 549-557.

Lorenz, E. N.[2006b] "An attractor embedded in the atmosphere," *Tellus,* **58A**, *425–429.*

Morales C. A., Pacifico M. J., & Pujals E. R. [2002] "Robust transitive singular sets for 3-flows are partially hyperbolic attractors or repellers," Reprint.

Palis J. [2000] "A global view of dynamics and a conjecture on the denseness of attractors," *Astérisque*, 261, 335-347.

Palis J. and de Melo W. [1982] "Geometric theory of dynamical systems," Springer-Verlag.

Phillips, O. M. [1960] "On the dynamics of unsteady gravity waves of finite amplitude. Part I. *J. Fluid Mech*. Vol. 9, 193-217.

Poincare, H. [1892] "*Les Methods Nouvelles de la Mecanique Celeste*," Vol. I, Gauthier-Villars: Paris

Smale, S. [1998] Mathematical problems for the next century," *The Mathematical Intelligencer*, **20**, 7-15.

Snyder, H.A. [1969] "Wave-number selection at finite amplitude in rotating Couette flow," *Journal of Fluid Mechanics*, Vol. 35, 273-298.

Stokes G. G. [1847] *"On the Theory of Oscillatory Waves*," Cambridge, Transactions, Vol. 8, 441-473.

Teixeira, J., C. A. Reynolds, and K. Judd, [2007] "Time step sensitivity of nonlinear atmospheric models: numerical convergence, truncation error growth, and ensemble design." *J. Atmos. Sci.*, **64**, 175-189.

Viana, Marcelo [2000] "What's new on Lorenz strange attractors?" *The Mathematical Intelligencer*, **22**, 6-19.

Yao, L. S. [1999] "A resonant wave theory," *J. Fluid Mech.*, **395**, 237-251.

Yao, L. S. [2007]"Is a Direct Numerical Simulation of Chaos Possible? A Study of a Model Non-Linearity," *International Journal of Heat and Mass Transfer,* **50,** 2200–2207.

Yao, L. S. [2009] "Multiple solutions in fluid mechanics," *Nonlinear Analysis: Modeling and Control*, Vol. 14, No. 2, 263–279. [http://www.lana.lt/journal/issues.php]

Yao, L. S. [2010] "Computed Chaos or Numerical Errors," *Nonlinear Analysis: Modeling and Control*, Vol. 15, No. 1, 109-126. [http://www.lana.lt/journal/issues.php]

Yao, L.S. and Ghosh Moulic, Sandipan [1994] "Uncertainty of convection," *International Journal of Heat & Mass Transfer*, Vol. 37, pp.1713-1721.

Yao, L.S. and Ghosh Moulic, Sandipan [1995a] "Taylor-Couette instability with a continuous spectrum," *J. Appl. Mech.*, **62**, pp. 915-923.

Yao, L.S. and Ghosh Moulic, Sandipan [1995b] "Nonlinear Instability of Travelling Waves With a Continuous Spectrum," *International Journal of Heat & Mass Transfer*, Vol. 38, pp. 1751-1772.

Yao, L. S. & Hughes, D. [2008a] "Comment on "Time Step Sensitivity of Nonlinear Atmospheric Models: Numerical Convergence, Truncation Error Growth, and Ensemble Design" Teixeira et al. (2007)," *J. Atmos. Sci.*, **65**, No. 2, 681-682, February.

Yao, L. S. & Hughes, D. [2008b] "A comment on "Computational periodicity as observed in a simple system," by Edward N. Lorenz (2006a)," *Tellus **60A***, **4**, 803-805.